\documentclass[final] {aipproc}
\layoutstyle{6x9}

\begin{document}
\title{Detection of Cascades induced by Atmospheric Neutrinos in the 79-string IceCube Detector}
\classification{95.55.Vj, 14.60.Lm, 29.40.Ka, 95.85.Ry, 25.30.Pt}
\keywords      {DeepCore, IceCube, neutrino, cascades}

\author{Chang Hyon Ha for the IceCube Collaboration~\footnote{see the full author list 
    at~http://icecube.wisc.edu/collaboration/authors/current}}{
  address={Lawrence Berkeley National Laboratory, Berkeley, CA 94720, USA}
  ,altaddress={Dept. of Physics, University of California, Berkeley, CA 94720, USA}
}

\begin{abstract}
Neutrino production and oscillation physics can be studied by utilizing the very high flux 
of atmospheric neutrinos observed with IceCube. In a Cherenkov medium such as ice, atmospheric muon neutrino interactions 
create tracks while cascades (showers) are produced by atmospheric electron neutrinos and 
by neutral current interactions of all flavors. 
We present the first detection of atmospheric neutrino-induced cascades 
at energies between 30~GeV and 10~TeV
using the DeepCore array of the IceCube detector. 
Using 281 days of data, 1029 events are observed with 59\% predicted to be cascades. 
\end{abstract}

\maketitle

High energy neutrino telescopes, like the IceCube neutrino observatory are sensitive to all active flavors of neutrinos.
The detector, using the deep Antarctic ice sheet as a Cherenkov medium, observes many atmospheric $\nu_\mu$-induced muons 
through charged current (CC) interactions in the ice
at energies as high as 400~TeV~\cite{Diffuse-Warren-2011}.
However, searches for atmospheric $\nu_e$~CC, $\nu_\tau$~CC, and all-flavor neutral current (NC) interactions have 
so far only resulted in upper limits for the flux~\cite{Cascade-ICRC-2011,Cascade-Joanna-2011}. 
The main signature for these interactions is a compact shower, 
producing a roughly spherical light distribution, called a ``cascade.''
We report the observation of atmospheric cascades in IceCube, 
using the first data with the DeepCore low-energy extension at energies 
between 30~GeV and 10~TeV.
At these energies, the cascade detection channel is well suited for studying kaon production in air showers 
and detecting neutrino oscillation signatures such as $\nu_\mu \rightarrow \nu_\tau$.

DeepCore takes advantage of compact sensor spacing, 
high quantum efficiency photomultiplier tubes, 
deployment in the clearest ice,
and a lower trigger threshold than 
the surrounding IceCube detector to observe neutrinos as low as 10~GeV~\cite{DCDesign-Doug-2012}.
In this analysis, we consider the data collected between May 31, 2010 and May 13, 2011.
The IceCube detector consisted of 79 strings where the DeepCore subarray was defined as
six densely instrumented strings optimized for low energies~\cite{DCDesign-Doug-2012}
and the seven adjacent standard strings.
This fiducial volume contains 454 digital optical modules (DOMs) deployed at depths below 2100~m.
The raw data, after applying quality criteria (90\% of data retained),
is composed of light signals (``hits'') in the DOMs. 
The hits are collected in two modes - with and without a local coincidence (LC) requirement on signals in neighboring DOMs.
Details of the LC logic can be found in Ref.~\cite{DOM-Matis-2008}.
With cuts designed to remove noise hits, 
the extra non-LC hits enhance reconstruction, background rejection, and particle identification especially 
for the lowest energy events.

A low threshold trigger, requiring minimum three LC hits 
within a time window of 2500~ns, is applied in the fiducial region.
A special filter is run on the triggered event sample at the South Pole to reject
cosmic-ray muons which penetrate the fiducial region, resulting in a passing rate of 17.5~Hz. 
A factor of 10 reduction in data compared to the triggered events 
is achieved by this algorithm 
while maintaining 99\% of the atmospheric neutrinos that interact in the fiducial volume.
The filtered data are sent north for further processing.
The details of the trigger and filter algorithms are in Ref.~\cite{DCDesign-Doug-2012}.

Atmospheric neutrinos are the decay products of 
charged pions, kaons, and muons created in cosmic-ray interactions
with nucleons in the atmosphere. 
At high energies, above $\sim$100~GeV, atmospheric $\nu_e$ are produced
mainly by semileptonic kaon decays like $K^+~\rightarrow~\pi^0e^+\nu_e$ mode.
The muons from meson decays with higher energies are more likely to interact before they decay, 
not producing $\nu_e$ in the atmosphere.
Thus, $\nu_e$ production is highly suppressed compared to $\nu_\mu$~\cite{Kaon-Agrawal-1996}.
The theoretical uncertainties associated with the $\nu_e$ flux are large 
due to the model uncertainties in the kaon production and the lack of $\nu_e$ measurements 
at these energies~\cite{BartolError-2006,IntHonda-2007}.

The backgrounds for the atmospheric $\nu$-induced cascades consist of  
cosmic-ray muons that mimic signal events
and $\nu_{\mu}$~CC events with low energy muons.
The cascade analysis identifies the topology of the light pattern 
and enforces containment of the signal to reject those backgrounds.
Veto techniques in the DeepCore fiducial volume
remove more than six orders of magnitude 
of the background events while retaining reasonable efficiency 
for atmospheric $\nu$-induced cascades~\cite{Ha-Thesis-2011}.
Following the veto step, the background rejection is performed in three stages,
each with a Boosted Decision Tree (BDT) \cite{BDT-Hocker-2007}.
The first stage is formed from five variables that characterize cascade signals in the detector
and are built from the spherical hit pattern, 
localized time structure, and quick charge deposition.
This selection reduces the data rate to 0.1~Hz, a factor of $\sim$1800 with respect to the trigger.
The simulation predicts the atmospheric $\nu_{e}$ rate at this stage to be $6.2 \times 10^{-4}$~Hz,
corresponding to 63\% efficiency with respect to the trigger.
In the next stage, we carry out more sophisticated likelihood reconstructions on the reduced dataset, utilizing
the scattering and absorption of Cherenkov photons in the ice~\cite{AMANDA-Christopher-2004}. 
From these reconstruction results, a seven variable BDT is formed and its selection
reduces the atmospheric muon background 
to $5.0 \times 10^{-4}$~Hz, rejecting an additional factor
of 200 ($3.6 \times 10^5$ cumulatively) 
and retaining $\sim$40\% of the $\nu_e$ 
signal ($2.6 \times 10^{-4}$ Hz) compared to the first stage.
At the final stage, tight cuts are made on the previous sample which contains
a large fraction of atmospheric neutrinos.
The cuts aim for high purity cascade detection by rejecting
$\nu_{\mu}$~CC events where possible.
The background rejection techniques are discussed in more detail in Refs.~\cite{Ha-Thesis-2011,TAUP-Ha-2011}.

\begin{figure}
  \vspace{10mm}
  \includegraphics[width=3.0in]{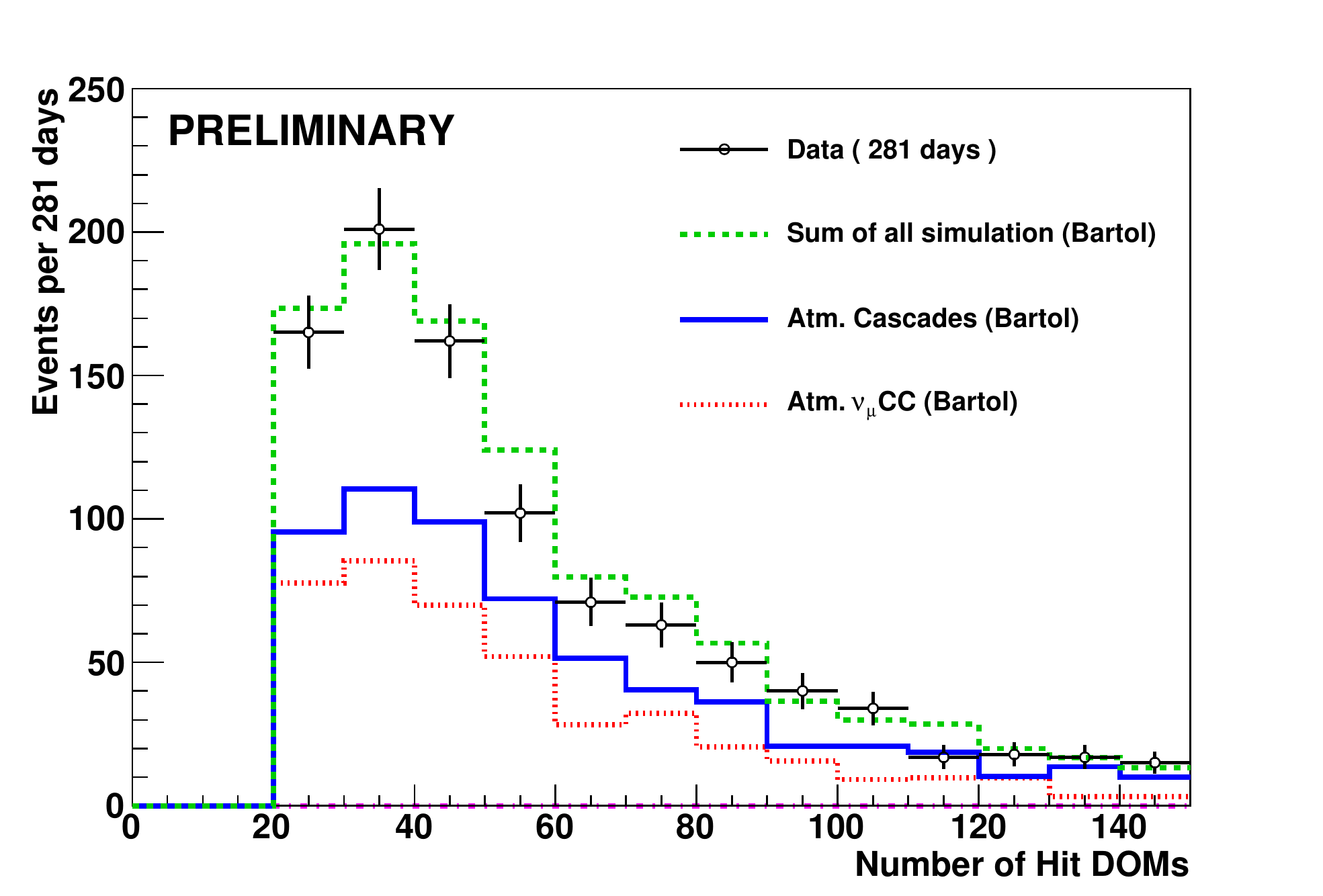}
  \includegraphics[width=2.8in]{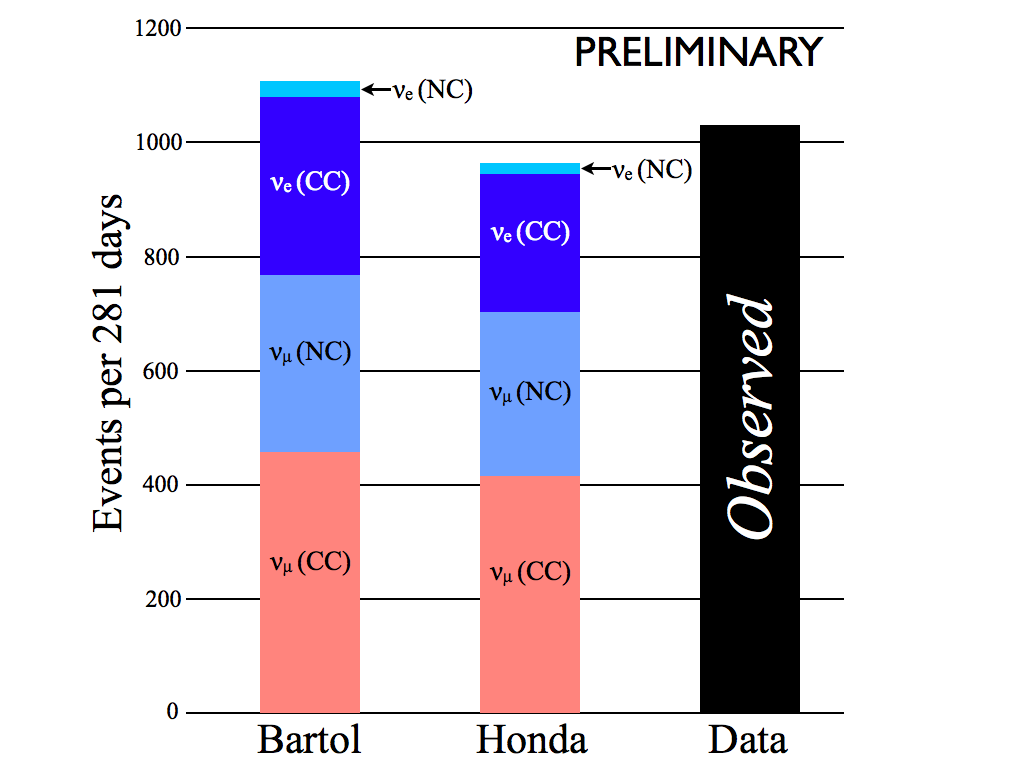}
  \caption{The event rate as a function of the number of hit DOMs (left). 
    The total simulated rate is consistent with the measurement from 281 days of data.
    Errors are statistical only.
    The bar histogram (right) shows simulation predictions for different
    interactions with two atmospheric flux models (Bartol~\cite{Bartol-2004} and Honda~\cite{Honda-2007})
    and the observed data rate.
  }
  \label{hard}
\end{figure}

We observe in total 1029 events in 281 days of data
and expect 651 (550) cascades and 455 (415) tracks from simulations using the
Bartol (Honda) atmospheric neutrino model~\cite{Bartol-2004, Honda-2007}.
The observation is consistent with either model, as shown in Fig.~\ref{hard}.
Approximately half of the cascades are predicted to be $\nu_{e}$ events and
the other half $\nu_{\mu}$~NC events.
The residual $\nu_{\mu}$~CC background events have short outgoing muons 
with simulations indicating
a median track length of 80~m 
where the tracks are not easily detected.
In the currently available 28 hours of atmospheric muon simulation, no events remain.
The 90\% upper limit on this prediction is 554 events in 281 days.
The simulated statistics are being increased.
Note that neutrino oscillations of $\nu_{\mu} \rightarrow \nu_{x}$ have a small ($<$3\%) 
effect due to relatively high energy 
of the cascades ($\rm <E_\nu>\sim$180~GeV).
The rate predicted by the Honda model is lower than that from the Bartol model
because the two models treat the normalization of the cosmic-ray spectrum and the kaon production in the 
atmosphere differently~\cite{Kaon-Agrawal-1996,IntHonda-2007}.
Systematic uncertainties are not included.

We have reported on the observation of atmospheric neutrino-induced cascade events in IceCube.
The observations are consistent with models of atmospheric neutrinos between approximately 30~GeV and 10~TeV.
Systematic errors arise from light sensitivity of DOMs, ice modeling, cosmic-ray composition, neutrino-nucleon cross section, 
and atmospheric neutrino flux model. 
The background contamination from atmospheric muons and systematic uncertaities are currently under evalution.
We expect results using a similar veto technique, 
aiming at neutrino oscillation measurements and WIMP dark matter searches in near future.
\bibliographystyle{aipproc} 
\bibliography{234_ha}
\end{document}